To appear in *Comprehensive Philosophy of Science* (Elsevier;
editor-in-chief Sven-Ove Hansson Biology section editor, Francesca Merlin

# Biology and Physics

Stuart A. Newman[1] and Sahotra Sarkar[2]

[1]Department of Cell and Molecular Physiology, New York Medical College, Valhalla, NY 10595, USA, Email: newman@nymc.edu

[2] Departments of Philosophy and Integrative Biology, University of Texas at Austin, Austin, TX 78712, USA, Email: sarkar@austin.utexas.edu

## Key Points

• Presents a "forms of matter" perspective on physical theory and posit biological physics as a subdiscipline of physics concerned with living matter.

• Distinguishes "unitary" and "nonunitary" forms of matter in relation to scaling of their inherent properties.

• Defines and provides examples of "generic," "biogeneric," and "nongeneric" forms of living matter.

• Relates biological physics to the past philosophical initiatives of dialectical materialism and logical empiricism and contrasts it to more recent "informationism."

## 1. Introduction

Juxtaposing the terms "biology" and "physics," as in the title of this article, immediately raises questions about the relationship between the disciplines and their objects of study. While their domains of applicability – the living and nonliving – are uncontroversial, since life has no universally accepted definition or an agreed-on scenario for its origin, a relationship of reductive inclusion is generally posited: viz, biology as an organizationally complex, derived manifestation of a more fundamental physical reality. Physics, then, has as its domain both living and nonliving phenomena.

When analyzed in greater detail, however, it is reasonable to infer that the items studied by biology such as microbes, animals, and plants, are different versions of one kind of thing, descendants of a putative common ancestor and therefore unified by evolution. The fundamental entities of physics, in contrast – fields and particles, mass and charge, liquids and solids, and so on – are ontologically distinct kinds.

This conclusion does not negate the fact that features of organisms can be, and are, studied using physical methods originally devised for nonliving systems. This practice, under the long-established rubric of *biophysics,* deals with questions like how nerves conduct electrical signals and how muscle cells generate mechanical force. But biophysics is not concerned with the kinds of questions ("first principles") such as how living systems are distinguished from nonliving ones, or how phenotypic traits relate to genotypes. This falls to the more recent initiative of the "physics of life," or *biological physics* (National Academies of Sciences, 2022), in which biology is treated as an authentic and distinct branch of physics, not just an object of it.

Notwithstanding mid-twentieth century philosophical initiatives to unify the sciences (Sarkar, 1998; 2025), it is now broadly acknowledged that the branches of physics are neither nested hierarchies of levels of organization nor recipes for reduction of one level to another. Thermodynamics is not explicable by quantum electrodynamics (or the reverse), nor does planetary motion depend on chemistry. The discourse around biology and physics, however, often represents a throwback to the older idea that the macroscopic can be explained by the microscopic and thus that an understanding of anatomy, physiology, and behavior can thereby be found in the properties and interactions of molecules, chemical reactions, and electrochemical networks.

The thesis that such explanations will always eventually be forthcoming is called reductionism (or, more precisely, strong or parts-whole reductionisms) (Sarkar 1998). Its failure within physics, which was recognized by the mid-twentieth century (Shimony, 1987; Batterman, 2021), did not prevent the new discipline of molecular biology, especially owing to the influence of Linus Pauling, from systematically pursuing reductions (that is, parts-whole explanations) starting in the late 1940s (Sarkar 1998). In many areas of cellular and subcellular biology, reductions were easily found and taken to be important, for instance, in the explanation of cooperative phenomena such as allostery or feedback regulation  such as that seen in the operon model. However, reductionism has never had much success at larger spatial scales, for instance, in the morphological development of complex organisms. These domains have, instead been hospitable territory for *organicism* with its emphasis on the role of biologically organized complexes influencing the organization and behavior of parts (and the related doctrines of emergentism and holism) (Gilbert and Sarkar, 2000; Nicholson and Gawne 2015). We will return to the issues of reductionism, organicism, emergentism, and holism in the concluding Section 7.

Meanwhile, our intention in this article is to dispel the reductionist view of living matter, though not by embracing doctrines traditionally associated with organicism and emergentism though we are sympathetic to a dialectical organicism (see Section 7). Instead, we characterize the distinct *forms of matter* (a notion that we explicate in detail below) that constitute cells and organisms, and the physics (old, new, and dimly understood) that organizes them. We adopt the framework of *materialism*, the position that everything that exists can be viewed as a manifestation of matter. This perspective has several sources. Regarding ontological claims the most important of these is *dialectical materialism* as advanced by Friedrich Engels in *Anti-Dühring* (1877-78) (Engels, 1987a) and the posthumous *Dialectics of Nature* (1872-82) (Engels, 1987b). (Many of the British organicists of the mid-twentieth century identified with this approach; Gilbert and Sarkar, 2000; Nicholson and Gawne, 2015; as did the ecological and population biologists Levins and Lewontin [1985]). If we restrict attention to epistemological claims, the relevant source is the physicalism of the logical empiricists, Otto Neurath, who was also an Austro-Marxist, and Rudolf Carnap, both of whom explicitly identified physicalism with materialism (Sarkar 2025, Chapter 5). (However, we do not use “physicalism” because it may generate ambiguity insofar as contemporary philosophers of mind interpret the term very differently (e.g. Strawson and Freeman 2024).)

Dialectical materialism was Engels’ generalization of *historical materialism*, Marx’s and Engels’ theory of social formations. The extension to the natural sciences was spurred by the identification by nineteenth-century scientists of forms of matter beyond the one that Newton described in his mechanics, the laws he had formulated that were the basis for Laplace’s claim in 1814 that they provided a “theory of everything” (Laplace [1840], 1951, 4), i.e., chemistry, optics, thermodynamics, and electrodynamics.

Engels emphasized both the qualitative distinctiveness of all such forms of matter and their continuity, as evidenced by their capacities to interact with one another (e.g., heat speeds up many chemical reactions, moving magnets generate electrical current and currents generate magnetic fields). In his words: “The whole of nature accessible to us forms a system, an interconnected totality of bodies, and by bodies we understand here all material existences extending from stars to atoms…” (Engels, 1872, 363). The physicist and historian Boris Hessen, in his 1931 essay on Newton and his successors, characterized the departure represented by Engels’ materialist synthesis as a major epistemological break in the philosophy of science (Hessen et al., 2009.)

Writing a century after Engels, the Marxist philosopher Louis Althusser (1969) reiterated this view:

> …the philosophical *category* of matter, which is conjointly a thesis of existence and a thesis of *objectivity*, can never be confused with the contents of the scientific *concepts* of matter. The scientific concepts of matter define knowledges, relative to the historical state of the sciences, about the objects of those sciences. The content of the scientific concept of matter changes with the development, i.e., with the deepening of scientific knowledge. The meaning of the philosophical category of matter does not change, since it does not apply to any object of science, but affirms the *objectivity* of all scientific knowledge of an object. (p. 49)

Dialectical materialism had its roots in the idealist philosophy of Hegel rather than the mechanized world picture from the seventeenths and eighteenth centuries (Dijksterhuis, 1986). In contrast to Aristotle’s hylomorphism, in which qualitatively indifferent matter is infused with one or another formative principle, in Engels’ view there is no “primitive” matter, but each kind has an inherent “motion” that characterizes its activity and capacity to interact with other forms (Bazhenov, 1975). It is therefore unsurprising that dialectical materialism has always aspired to comprehend – rather than deflate or reduce to known physics and chemistry (e.g., as in “eliminative materialism” Churchland, 1981) – phenomena such as cognition and consciousness (Vygotsky, 1980; Rubinstein, 1946). Here we do not pursue the possible bases in novel forms of matter of these mentalistic living processes but confine our discussion to the morphological and physiological functions of multicellular organisms. For morphology, according to our perspective, these depend on the collective physical properties of their constituent cells and, for physiology, on the appropriation and partitioning in specialized tissues of living processes intrinsic to cellular life. Cells themselves by all evidence conform to known physical principles (however see the discussion in Section 6 of "biomolecular condensates") but their origins and internal organizational properties are not understood from first principles, and these aspects will only be touched on briefly in this article.

Our discussion repeatedly invokes different "forms of matter." We do not try to provide an explicit definition of this concept or provide rigorous criteria for its use but apply it variously to distinct states of a single substance on different sides of a phase transition (e.g., ice and liquid water; magnetized and unmagnetized iron), composites with apparently novel features assembled from other units (e.g., atoms from nucleons and electrons; proteins from amino acids) and so on. As these examples show, the basis of the novel qualities can be relatively transparent or more obscure. In some cases, there are physical theories for the constitution of new forms (e.g., the hydrogen atom) and in others these are lacking (e.g., biomolecular condensates). In all cases, however, forms of matter are characterized by dispositions or *inherencies* that distinguish them from other such forms; these are dispositional because they are expressed in a conditional fashion. For example, the capacity to form waves or vortices is inherent to liquids but these dispositions are not always realized, and rarely realized at the same time. These considerations apply both at the level of structure and process and the two are entangled; from this entanglement we find new forms of matter.

In what follows we describe several forms of living matter that arise from the association of cellular subunits and discuss how the inherencies and transformational propensities of these materials determine organismal forms and functions. To begin with, however, we provide a sketch of thermodynamics, which specifies relationships between heat, work, and tendency to spontaneous change, which due to its generality is used, and sometimes misused, in explanations of living processes. We also present a typology of matter that will facilitate comparison between nonliving and living matter and the unique aspects of the latter that compel the pursuit of new principles.

## 2. Thermodynamics and forms of matter

In Einstein's view, thermodynamics is "the only physical theory of universal content concerning which I am convinced that, within the framework of the applicability of its basic concepts, it will never be overthrown" (Einstein 1949, p. 33). The domain of applicability of classical thermodynamics consists of continuous substances characterized by measurable properties like temperature, pressure, volume, and energy. When the theory was originally formulated in the first half of the nineteenth century the modern concepts of atoms and molecules were unknown, so there was no recognition that gases, solids and liquids – the forms of matter it dealt with – were actually composed of identical discrete subunits, and therefore discontinuous. In practice, however, *unitary* forms of matter, materials which, on scales larger than their subunits, have characteristic intensive properties (e.g., temperature, viscosity, elasticity) that are not found in the subunits themselves, can be treated as effectively continuous and are suitable objects of thermodynamic analysis. (Unitary *chemical* systems are ones that are well-mixed at the molecular level.) Unitary materials inhabit abstract morphospaces in which their *inherencies* (Newman, 2019b, 2024) – that is, propensity to exhibit characteristic organized structures (e.g., ripples or vortices in liquid water, crystals, plates, or nanotubes in solid carbon) – are realized under specific conditions. Reorganization can occur when the material is part of an *open system* and mechanical work is performed on it or it is subject to a flux of energy or matter (or both) through it. The distinct "morphotypes" can be maintained when the system is at mechanical rest (as with carbon allotropes) or when it is thermodynamically open and dynamically displaced from equilibrium. Prigogine called such nonequilibrium forms and patterns "dissipative structures" (Goldbeter, 2018; Prigogine, 1969).

Biological systems are almost invariably thermodynamically open, but there are exceptions such as the spores of bacteria and yeast (Walker et al., 2024), and the desiccated cysts of brine shrimp (Hengherr et al., 2011), in which there is no exchange with the environment and active processes are suspended indefinitely (or nearly so), but the potential for revival is maintained structurally. There are also cases like mixtures of different embryonic cell types (discussed below) where despite the cells' dependence on energy and matter fluxes to remain viable, the aggregate undergoes demixing and phase separation as if were a closed system of two mixed liquids (Steinberg and Poole, 1982).

Matter and energy fluxes can also drive chemical reactions among the constituents of an open system, leading to compositional changes and, potentially, temporal or spatial chemical nonuniformities. This is exemplified in nonliving systems by the *oscillatory* Briggs-Rauscher reaction, in which the redox state of a stirred system changes abruptly and cyclically every ~15 seconds (Dulos and De Kepper, 1983) and the *reaction-diffusion* systems described by Turing (1952) in which *spatially periodic* patterns of stripes, spots, and spirals of chemical composition with ~0.2 mm repeats are demonstrable in planar media (Castets et al., 1990). Another potentiality of coupled chemical reactions in thermodynamically open systems is *multistability*, the existence of more than one stationary (unchanging in time) compositional state, each surrounded by a "basin of attraction" in the system's state space. Discrete transitions between attractor states can occur if the system is perturbed (reviewed in Forgacs and Newman (2005)). The mobilization of stored energy, an attribute that is rare but not unknown in nonliving systems, defines some of these materials as "excitable media" (Levine and Ben-Jacob, 2004). Further, the fact that their living-cell subunits can exhibit motility makes animal tissues novel forms of "active matter" with properties not readily predicted by physical laws formulated for conventional viscoelastic materials (Bernheim-Groswasser et al., 2018; Gross et al., 2017).

Each of these behaviors or outcomes in thermodynamically open, nonequilibrium systems (collectively termed "self-organization") has been invoked to account for biological phenomena – temporal oscillations for physiological activities such as circadian rhythms or developmental ones like the formation of body segments, spatial periodicities (also known as standing waves) for development of the digits of the tetrapod limb and for hair and feather follicles of skin, and multistability for transitions between cell states during development and for the panoply of cell types generated from an organism's genome. These models have had some success, but in certain cases have led to serious misconceptions, as we discuss in Section 4 (see also Newman, 2022b).

The simplest and most direct way that physics can account for a biological phenomenon is if it is "generic," that is, the principles and effects invoked pertain to both living and nonliving matter (Newman and Comper, 1990). Gravity is an obvious example, since it affects, in the same way, anything that has mass. Molecular diffusion, mediated by Brownian motion, is another example, establishing chemical gradients in embryos from which developing cells can take their cues. Living entities can also follow Fick's diffusion law on a larger spatial scale, as with the randomly oriented movement of cells in a tissue mass. Here it is not Brownian motion that drives diffusion, but locomotion associated with fluctuations in cell shape. Such effects and their outcomes (e.g., the rounding up of a cell aggregate by a cell-scale surface tension effect) have been termed "biogeneric" (Newman, 2016). This term is applied when the subunit behaviors are due to active living processes (e.g., nondirectional cytoskeletal contractility; secretion of a neighboring cell-altering protein), but the effect on the material in which it occurs is formally similar enough to nonliving subunit processes (e.g., random molecular collisions; formation of chemically reactive product) to justify employing generic-type models (e.g., Glazier and Graner, 1993; Hentschel et al., 2004). As we describe below, not all cell-cell behaviors and interactions constitute biogeneric phenomena, and their physical modeling requires additional layers of complexity.

In many cases, the physical nature of a developmental process has been transformed subsequent to its evolutionary origination due to selective or non-selective regimens (Müller and Newman, 1999; Newman, 2019b; True and Haag, 2001). Here a "generic" appearance can be deceptive. The seven evenly spaced uniformly wide stripes of the even-skipped transcription factor in the egg cytoplasm of *Drosophila* look like the result of a Turing process (and may have originated as such (Newman, 1993; Salazar-Ciudad et al., 2001), but in fact is a product of the operation of hierarchically regulated genetic determinants of segmentation (Akam, 1989).

Developmental mechanisms that are "non-generic" (i.e., unlike formative processes found in the non-living world) do not only result from the hodge-podge of hybrid mechanisms and so-called "kludges" (O'Malley, 2011) that may accumulate gradually over evolution. Some biomaterials, such as the developing tissues of embryophytes (land plants), were non-generic from the start (see Section 5). Though subject to physical analysis and mathematical modeling, they are unlike any non-living materials (Niklas and Spatz, 2012) (see Section 5).

Most applications of the physics of materials to living systems pertain to forms of matter for which *thermodynamic* concepts are relevant: quasi-continuous materials uniformly composed of cellular subunits, e.g., embryos and their primordia, aggregates of bacteria or social amoebae, tumor masses (Newman et al., 2025). Such materials are characterized by *state functions* like internal energy (U), enthalpy (H), entropy (S), and Gibbs free energy (G). State functions are so designated because they only depend on the current state of a system, not the path taken to get there. (Doubling the volume of a gas, for example, will always incur the same change of total energy, no matter how it occurs, but different ways of achieving it can involve adding different amounts of heat, so heat is not a state function.) In lieu of defining state functions in absolute terms, we list some of the ways they may change under typical biological conditions (constant temperature and pressure):

- Increase in total energy (ΔU) occurs when work is performed on, or matter added to the system.
- Increase in enthalpy (ΔH) occurs when heat is added to the system.
- Increase in entropy (ΔS) occurs when energy becomes dissipated and therefore less available to perform useful work. The fact that S never decreases in a closed system is the thermodynamic basis for (time) irreversibility.
- Gibbs free energy decreases ($\Delta G < 0$) when the system performs useful work; how much G is available depends, among other parameters, on the amount and concentration of U. Thus, if a process dissipates U, the amount of G available to perform it is less.

State functions are not readily definable for matter that is *non-unitary*, i.e., compositionally and spatially discontinuous or heterogeneous, and for which component subsystems are not in mutual equilibrium. This applies to assemblages like cities, ecosystems, or living cells, although each contains subsystems that are subject to thermodynamic analysis. In an attempt to characterize the thermodynamic nature of living systems, Kauffman (2000) wrote about "work-constraint cycles," in which evolved molecular processes such as "autocatalytic sets" of nucleic acids expended energy to build elements of the set responsible for performing this constructive activity. Mossio and Moreno (2015), in what they term the "organizational approach," extended this idea by proposing that living cells are constituted by a "closure of constraints" in which the products of interconnected systems of work cycles support one another's existence and activity. The organizational approach is thus an instantiation of the earlier "autopoiesis" model of living systems of Maturana and Varela (1972) in which a system of interacting biochemical processes were asserted to produce the conditions for their mutual production. Neither model contains an account of the origin of such arrangements, leaving in place Kant's conundrum of the organism as *natural purpose* (Kant [1790] 1966).

Finally, thermodynamic terminology has gained currency in many theoretical biology venues in the form "informational entropy" and Bayesian free energy minimization. This initiative derives from Boltzmann's (1872) effort, in his kinetic theory, to reconcile thermodynamics with the existence of atoms and Gibbs's (1902) interpretation of irreversibility in terms of uncertainty, order and disorder, and draws on Shannon's (1948) mathematical theory of communication. Such models can successfully enlist the mathematics of thermodynamics to predict the behavior of complex systems, but only when the possible realizations ("microstates") consistent with the system's macroscopic description can be characterized (e.g., (Harte et al., 2022; 2024)). This is often not the case, making the approach prone to the subjectivism and arbitrariness of "biological informationism" (Newman and Sarkar, in press; Sarkar, 1996, 1998).

In what follows, we consider the physics of unitary forms of biological matter in which "ontic" or material (i.e., in contrast to informational; Swenson, 2025a,b) thermodynamic functions can be defined, at least in principle. We distinguish whether the materials and processes we describe are (i) *generic* or *biogeneric* (as much of animal development appears to be), (ii) *nongeneric* (in the sense of having no corresponding form of matter in the nonliving world) but largely understandable in terms of known physics and chemistry (e.g., plant development), or (iii) *novel and enigmatic*, possibly requiring new physical theories of forms of matter peculiar to living systems, such as biomolecular condensates, employed, for example, in the metazoan developmental gene regulatory apparatus (reviewed in Newman, 2020).

Examples of the categories of matter discussed in this article are listed in Table 1.

## 3. Animal morphogenesis and pattern formation: the physics of biogeneric matter

The study of *morphogenesis* (shaping and reshaping of tissues) and *pattern formation* (arrangement of cell types) in animal embryos has been facilitated by the fact that many of the relevant tissues have liquid-like properties at early stages. As they develop, tissue masses undergo changes resembling those seen in nonliving viscoelastic liquids, although the subunits of the materials are cells instead of atoms, small molecules, or polymers. The tissues in question are therefore "biogeneric," and the default accounts of their behavior are physical theories and models that have been developed for, or in analogy to, nonliving liquids.

### *Liquid tissues and their inherencies*

The cells of early-stage animal embryos are attached to one another by "classical cadherins," members of a family of homophilic cell surface proteins with weak, reversible binding affinities (Halbleib and Nelson, 2006). This enables formation of cohesive masses in which cells collectively maintain mutual contacts although their nearest neighbors are continually changing. These are precisely the criteria of the liquid state (Widom, 1967). For nonliving liquids, the atomic or molecular subunits move randomly by Brownian motion while cohering by Van der Waals forces, hydrogen bonding, or polymer entanglement. For liquid tissues of metazoan embryos, random motion of the cellular subunits is effected by cytoplasmically driven fluctuations in cell shape, and cohesion by the aforementioned cadherins. The validity of the liquid-tissue concept is supported by the success of CompuCell3D, a computational framework based on the Cellular Potts model of the thermodynamics of biogeneric cell associations (Glazier and Graner, 1993), in simulating morphogenesis (of limbs, blood vessels, kidneys and other organs) and cancer (Izaguirre et al, 2004; Chaturvedi et al., 2005; Merks et al., 2006; Tripodi et al., 2010; Szabó and Merks, 2013; Tikka et al., 2022).

These developing tissues are thus not merely analogs of liquids; they are genuine liquid-state materials:

- Liquid tissues are typically incompressible and undergo *viscous flow* in a manner described by the Navier-Stokes equation (Iber et al., 2015).
- They exhibit *surface tension*, making the default shape of tissue masses composed of *spatially isotropic* cells (e.g., blastula-stage embryos) spherical.
- *Adhesively distinctive* cells will *sort out* when intermixed, and the sorted populations will undergo *phase separation* and form immiscible layers (reviewed in (Forgacs and Newman, 2005).
- Cells with *adhesively nonuniform surfaces* will spontaneously arrange into tissue masses with interior *cavities* or *lumens*, analogously to micelles forming in liquids composed of amphipathic molecules.
- Within liquid tissues, cells that are *elongated* rather than isotropic can interdigitate similarly to the subunits of *liquid crystals*, reshaping the mass, which typically lengthens perpendicularly to the cells' long axes (termed *convergent extension*).

In general, animal embryos start out spherically, form layers and interior spaces, and then elongate as they develop into the characteristic bodies of members of their respective species. The role of liquid-tissue physics, exemplified by each of the processes listed above can, be discerned experimentally in these transformations (reviewed in Forgacs and Newman, 2005). This, the success of simulations of morphogenesis by models such as those generated within the CompuCell3D framework, and examination of the fossil record and inferred phylogenesis of the animals has led to the proposal that the main features of metazoan body plans arose via inherencies of the liquid tissue morphospace (Newman, 1994, 2016) rather than the gradualism of conventional evolutionary theory (Budd and Jensen, 2017).

*Phase behaviors and transformations of liquid tissues*

The biogeneric (not simply generic) character of the relevant materials adds important caveats and additions to this picture of liquid tissues when it comes to understanding the physical mechanisms of development of present-day species and their evolution. The liquid properties of developing animal tissues pertain at scales at which these materials are unitary, i.e., when the individuality of the subunits can be ignored. But the subunits in these cases are cells, not atoms or molecules. They are loci of mechanical, biosynthetic, and electrical activity, and can shape-shift in response to their local environment, all of which complicates the assumption of generic physics. Further, cells are the venues of genes, the differential expression or mutation of which can change subunit properties over developmental and evolutionary time, respectively.

The mechanical activity of cells, for instance, can change the physical character of tissues in ways that are unknown in molecular liquids. As noted above, for example, the shape of a tissue mass will differ if the cells' shapes are isotropic or non-isotropic. Through induced rearrangements of their cytoskeletal fibers, cells can actively vary their shapes, thereby transforming an amorphous liquid tissue into a liquid crystalline state with consequent reshaping. The same kinds of interior rearrangements that lead to cell elongation can route adhesive molecules to distinct regions of the cell surface, inducing lumen formation. They can also cause cells to contract, leading in certain tissue primordia to multicellular *condensations*, which template skeletal structures such as those of backbones and limbs of vertebrate organisms. The mechanical transformation of subunits is a fundamental departure of biogeneric liquids from generic ones and an important basis of embryogenesis (Newman, 2019b; Newman and Bhat, 2008, 2009).

The synthetic capabilities of cells can also modify the physical properties of liquid tissues, causing them in some cases to no longer fit that description, behaving instead as elastic sheets or solids. Certain (i.e., epithelial) cells produce, secrete the precursors and provide conditions for the assembly of flexible, planar extracellular *basal lamina* on which they attach and spread. All tissue folds, including those that generate appendages and glands, result from this effect. Other (e.g., connective tissue) cell types secrete and surround themselves with *interstitial extracellular matrices* (ECMs) that define the viscoelastic properties of the respective tissues or even solidify them in the cases of bones and teeth. Solidification in this context may arise from the familiar first-order phase transition of crystallization-based mineralization, or a biology-specific second-order phase transition of assembling fibers of ECM protein collagen (Forgacs et al., 2003).

In addition to components of the ECM, cells of developing animal tissues produce and secrete molecules (proteins and nonproteins) that influence the fate of their immediate or more distant neighbors. Many of the key gene products involved in development are highly conserved across the metazoans and are referred to as the evo-devo (i.e., evolutionary-developmental) “toolkit” (Carroll, 2005). These include locally acting factors (e.g., Wnt) that alter cell surface or cell shape polarity (as described above) and the Notch-Delta cell surface ligand-receptor pair that mediates *lateral inhibition* by which cells cause their immediate neighbors to function differently from themselves. Also included in this class are *morphogens* (e.g., BMP, FGF), secreted proteins that spread from their cellular source. This transport function can occur generically, i.e., by passive molecular diffusion through the ECM. or biogenerically, e.g., by one or another energy-consuming process in which the cells along the way can facilitate or boost the morphogen’s transit. In such cases, gradients will form that can affect the behavior or states of distant cells in a concentration-dependent fashion (Kicheva and Briscoe, 2023).

*Dynamical patterning modules and the emergence of tissue heterogeneity*

The molecules of the evo-devo toolkit that function by mobilizing well-described physical effects (e.g., adhesion, diffusion, subunit polarization, lateral inhibition) have important roles in inducing transformations in the bulk properties of liquid tissues (as described above) or introducing morphological novelties consistent with tissue inherencies. The toolkit molecule-physics functionalities that mediate morphological development have been termed “dynamical patterning modules” (DPMs; Newman and Bhat, 2008, 2009).

The described lateral inhibition module (based on the Notch-Delta ligand-receptor pair), for example, can interact with other tissue functionalities to generate self-organizing instabilities and broken symmetries, employing oscillations of transcriptional regulators along the developing body axis of vertebrates. Weak coupling among the cells brings their oscillators into synchrony (Isomura and Kageyama, 2025; Strogatz, 2003). Such synchronization may provide a physical basis for the global coordination across embryonic tissues classically known as “morphogenetic fields” (Gilbert and Sarkar, 2000; Newman et al., 2021).

Lateral inhibition can also interact with morphogens in regulatory circuits that exhibit *reaction-diffusion instabilities* of the type described mathematically by Turing (1952). Such processes, or ones that also incorporate cell adhesion (Glimm et al., 2014), can generate standing waves of molecular determinants of repetitively arranged structures such as the digits of tetrapod limbs, or the feather or hair follicles of avian or mammalian skin. Addition to the Turing-type dynamics of mechanotransduction among a tissue’s cells (Farge, 2011) or advection of its secreted components (Recho et al., 2019) increases the variety of embryonic structures that may form, as does the simultaneous operation of biochemical patterning and three-dimensional tissue reshaping (Salazar-Ciudad et al., 2003). By these hybrid developmental processes, initially unitary or quasi-unitary materials (see Section 2) are converted into the architecturally heterogenous ones of mature animal bodies and organs.

Essential to the production of definitive animal bodies is the *differentiation* of specialized *cell types* (Barresi and Gilbert, 2024). These types, which range in number from fewer than 20 in the morphologically simplest animals (placozoans, poriferans) to as many as 300 in humans, implement physiological functions (mobility, digestion, sensation, etc.), and are induced to appear during development by complexions of molecular and mechanical determinants generated by the processes described above. An organism’s cell types involve differential expression of thousands of genes in partially overlapping but discrete patterns, with some analogy to the attractors of unitary dynamical systems like Boolean networks or systems of ordinary differential equations. Consequently, for many years it was widely accepted that the structure of the “developmental genome” of animals constituted such a dynamical system. Although there is evidence for dynamical switching behavior in transitions between cell states that are adjacent to each other in developmental lineages (Corson and Siggia, 2017; Glimm et al., 2023) there is now good reason to reject the idea that an organism’s cell types are the attractors of a single dynamical system (reviewed in (Newman, 2020a) and that “global attractor” model will not be discussed further here. A newer view of cell differentiation, involving the still obscure physical properties of biomolecular condensates, will be described below, in Section 6.

## 4. Developmental system drift obscures the roles of physical processes

The morphospace of liquid tissues along with transformations induced by ECMs and dynamical patterning modules are the basis for metazoan organisms exhibiting their characteristic body plans and organ forms. Natural selection may favor the propagation of certain morphological variants, but these must be consistent with the inherencies of multicellular matter, which are ultimately responsible for the major features of developmental and adult anatomy.

By determining their possible shapes and forms, the materiality of developing tissues limits the capacity of organisms to evolve morphologically. Morphological stasis is further reinforced by intra-organism integration of physiology and anatomy and ecologically based organism-organism and organism-affordance (in Gibson's sense of action opportunities; Gibson, 1966) interactions. While mutations that disrupt the morphological phenotype will generally be selected against, rewirings of the mechanisms that bring it about will often not be. Any alterations, for example, that preserve the morphology but generate it in a more robust or less costly fashion while preserving reproductive compatibility with populational cohorts will be at a premium. In addition, mutations that change developmental mechanisms in an indifferent fashion could persevere. If they produce ecologically adapted but reproductively insular subpopulations, they could wind up engendering new lineages. Such effects been called "developmental system drift" (DSD; True and Haag, 2001: "an evolutionary phenomenon whereby the genetic underpinnings of a trait in a common ancestor diverge in descendant lineages even as the trait itself remains conserved.") The conservation of form in the face of these mechanistic changes is referred to as "autonomization" of constructional elements (Müller and Newman, 1999) or phylogenetic "homomorphy" (Newman, 2019a). As described below, DSD can lead to substitutions of not only the genes and gene product interactions involved in bringing about a given form, but also of the physical processes underlying these morphological outcomes.

These phenomena are problematic for attempts to assimilate biology to physics due to the "palimpsestic" (by analogy to successively overwritten medieval parchments) effects of DSD. They obscure the likely originating processes underlying animal body plans and organ forms, making experimental evidence for the complexity of developmental mechanisms of present-day organisms ostensible refutations of biogeneric determination of form.

As an example, we can consider the long-held conviction that *differential interfacial tension*, initiated the formation of the layered gastrula stage. This proposal was first advanced by Steinberg (1970) as the "differential adhesion hypothesis," which focused strictly on the adhesive affinities of the cells' cadherins, and later made more fully biogeneric when *cortical tension* of the cellular cytoskeleton was recognized as also contributing to tissue interfacial free energy (Brodland, 2002; Manning et al., 2010). Direct measurement of single cell adhesive properties and cortical tensions predicted sorting behavior of intermixed zebrafish embryo cells and spatial arrangements of the resulting tissue masses (Krieg et al., 2008). Further, the physical theory of *wetting* provided an accurate account of the changes of the contact angle at the leading edge of the spreading epiblast (enveloping germ layer) of the zebrafish embryo over the yolk cell mass (enveloped germ layer (Wallmeyer et al., 2018).

Surprisingly, however, in vivo measurement of differences in germ layer tissue surface tension in the gastrulating embryo found the magnitude of this difference to be insufficient to drive progenitor cell sorting and tissue rearrangement under physiological conditions (Krens et al., 2017). Further experiments led to the attribution of the earlier in vitro results to disparities in the osmolarity of the embryo's interstitial fluid and the tissue culture medium employed (Krieg et al., 2008). The investigators suggested that rather than biogeneric interfacial forces, *directed cell movement* (a nongeneric, specifically biological, effect) was responsible for driving germ layer rearrangement in vivo (Krens et al., 2017). Although there is no evidence that permits us to decide whether interfacial tension or directed cell migration was the evolutionarily primitive mechanism for the establishment of germ layer boundaries and arrangement in vertebrates, the fact that the interfacial contours are predicted by the physics of wetting (a condition not automatically achieved by the directed migration mechanism) makes it plausible that the biogeneric liquid properties of animal tissues provided the evolutionary template that agential cell biological processes built upon in present-day zebrafish development.

**5. Plant development: the physics of nongeneric unitary biological matter**

Unlike the tissues of animals which can be analyzed with established physical models by virtue of their biogeneric liquid and viscoelastic properties, the tissues of multicellular plants and fungi are examples of nongeneric biomaterials. Being composed of identical subunits, both are unitary forms of matter. Plant tissues are solids, but deformable ones, since their extensible solid cell walls surround expandable cytosolic interiors. Such “cellular solids” (Niklas, 1992) have some properties analogous to foams (Weaire and Hutzler, 2000) but for the most part they have no natural nonliving counterparts.

Plant tissues are highly dynamic, more so in some respects than animal tissues. This is due to their unique mechanical properties and signaling modalities. For the first, an increase or decrease in turgor, induced, for example, by altered physiological conditions, place cell walls under increased or decreased tension. Higher stress activates proteins like expansins, which loosen the cell walls by disrupting hydrogen bonds between cellulose microfibrils and matrix polysaccharides (e.g., xyloglucans; pectins). As a consequence of slippage of these polymers the cell walls extend irreversibly. The direction of turgor-induced stress influences the orientation of new cellulose deposition by cellulose synthase complexes and the accessibility of matrix components to hydrolytic enzymes. Turgor variations also trigger the influx of calcium which modulates activity of cell wall synthesis enzymes (Niklas and Spatz, 2012).

Regarding signaling, *auxins*, hormones which (among other developmental effects) induce growth and cell shape change in stems, root formation, and light and gravity tropisms, are conveyed rapidly throughout the plant by membrane transporters but are subject to spatial restriction in their actions, such as at sites of leaf primordium initiation. Such patterning is dependent on *plasmodesmata* (PD), microscopic cytoplasmic channels that connect neighboring cells to each other, facilitating the local transport of small molecules like ions, sugars, auxins and other hormones, and transcription factors. Some of these factors regulate the aperture size of PD and therefore their permeability to auxins, and others affect the cell responsivity to auxins. Auxins, for their part, regulate the permeability of PD, leading to feedback effects with morphogenetic outcomes (Mellor et al., 2020; Niklas and Kutschera, 2012; Winnicki et al., 2021).

*Dynamical patterning modules of plants*

As with animal tissues, those of plants have characteristic morphospaces with associated dynamical patterning modules (DPMs) which reflect their morphogenetic propensities (Hernández-Hernández et al. (2012). In analogy to animal systems, combinations of these DPMs implement biochemical and mechanochemical reaction–diffusion mechanisms (Benítez et al., 2011). Models based on these effects can go some distance in emulating developmental phenomena such as *phyllotaxis*, the regular arrangement of lateral organs such as leaves and flowers around the plant stem, but a complete explanation appears to require lateral inhibitory interactions whose biological basis remains unknown (Bhatia and Heisler, 2018; Turing and Saunders, 1992). Indications that this "nongeneric" outcome (in the terminology introduced above) will eventually yield to a physical account is supported by a remarkable set of experiments conducted by Douady and Couder (1992), who devised a system in which magnetic fluid droplets were successively released from a stationary pipette. The droplets' mutual repulsion imparted an angular bias to their arrangement, that led to periodic spiral patterns conforming to the mathematical Fibonacci series exhibited, for example, by the disk florets on a sunflower head.

The foregoing description of the physical properties of the tissues of land plants contains the implicit suggestion that at some point in evolution a collection of preexisting living subunits (i.e., streptophyte algae) came together and produced a form of matter with novel nongeneric physical properties and morphological inherencies. The standard population biological narrative would predict that such a major transition would be a prolonged, incremental affair. It is therefore relevant to mention the recent report of an entirely different kind of biological solid with subunits that are not cells, but living starfish embryos. Here, an experimentally constructed unitary material was found to exhibit unprecedented rheological and self-organizing properties (Chao et al., 2026; Prakash, 2026). Examples like this add plausibility to the conjecture that novelties of form and function during evolution of multicellularity (e.g., in the animal and plant systems discussed) can arise by direct physical means rather than gradual selection.

## 6. Biomolecular condensates: novel forms of matter specific to living systems

With the discovery of biomolecular condensates (BMCs), we seem to be looking at a new form or class of matter, similarly to those scientists of the past who confronted (some under protest) purportedly unprecedented principles of heat and energy, electricity and magnetism, quantum mechanics, and relativity (Newman and Sarkar, in press). As mentioned in section 1, BMCs are a kind of material composed of known categories of ingredients with properties that are entirely unanticipated by the physics of those ingredients.

Biomolecular condensates are found throughout the cytoplasm and nucleus of eukaryotic cells (with examples as well in prokaryotes), typically presenting as "membraneless organelles." They form by processes akin to liquid-liquid phase separation, although whether they are truly liquid-state materials is unclear. The best studied BMCs, such as cytoplasmic stress granules and the nucleolus in the nucleus, consist of proteins with intrinsically disordered domains and, in many examples, RNA molecules. Along with fluid-like demixing of molecular component with very general sequence requirements, BMC assembly may include ordered cooperative assembly of macromolecules driven by allosteric regulation, site-specific interactions between conserved high-affinity binding partners, and percolation of polymers (i.e., connectivity-based second-order phase transitions). Energy-consuming processes such as ATP-driven cytoplasmic contractility and chemical reactions also participate in formation of some BMCs (Sanfeliu-Cerdán and Krieg, 2025). No other materials constituting or produced by cells, and certainly none identified in the nonliving world, have this order of complexity.

Biomolecular condensates are venues of a wide variety of cellular functions both in the cytoplasm (e.g., as organizers of cytoskeletal networks, signal transduction, metabolic compartmentalization, and sequestration of mRNA during cellular stress), and in the nucleus (e.g., in the ribosomal RNA-producing nucleolus and transcription) (Pei et al., 2025; Wiegand and Hyman, 2020). While they have been described as liquid-like, they are not composed of one or a few identical subunits that rearrange randomly, and in some cases (as in the nucleolus), they are layered and internally structured, but with dynamical properties unlike those of liquid crystals (Galvanetto et al., 2023; Quinodoz et al., 2025; Sanfeliu-Cerdán and Krieg, 2025). Importantly, *they are not unitary materials* like classic molecular liquids or even the embryonic liquid tissues described in Section 3, that is, they have no substructure above which their properties are uniform.

*Role of BMCs in cell differentiation*

Establishment of specialized cell and tissue types during embryogenesis in animals (with some parallels in plants) differs from physiological or "housekeeping" expression of individual genes associated with maintenance of the adult body (Arenas-Mena, 2017). This process of *cell differentiation* involves the concerted coregulation of suites of genes that underly the complex functions of the tens to hundreds of distinct cell types of a given species. All animals other than those of the morphologically simplest phylum, Placozoa, employ BMCs in a unique fashion in generating their cell types (reviewed in (Newman, 2020a,b; Pei et al., 2025).

Central to the mechanism by which BMCs coordinate differentiation are *enhancers* – hundreds to thousands of DNA elements scattered throughout the genome that amplify the transcription of target genes often located thousands of base pairs away. Their transcription products (eRNAs) promote condensate formation and recruit the multi-subunit scaffolding complex Mediator, transcription factors and other coactivators (Barral and Déjardin, 2023; Mattick, 2023). This results in deployment (in the form of specialized cell types) of functionally coordinated ancestrally co-expressed genes (Newman, 2022a).

Among the many functions in which BMCs are involved, cell differentiation best highlights the uniqueness of this form of matter, which appears to be a true interface between the RNA and protein worlds and possibly the living and nonliving. (Some theorists have suggested a key role for them in the origin of life; Hadarovich et al., 2025; Prosdocimi and Farias, 2025). Of all candidates for the mediation between nucleic acid sequence and physiological traits (one of the most persistent questions since the molecular gene was first posited (Sarkar 1998)), BMCs seem the most plausible, due to their extraordinary material properties, which, however, thus far defy physical theorizing. We conclude this section by describing these and how they differ from known forms of matter.

*BMCs as a novel nonequilibrium form of matter*

The forms of amorphous, condensed materials ("soft matter;" de Gennes, 1992)) analyzed by existing physical theories include liquids, gels, and glasses. The properties of liquids were described in Section 3. In gels, the subunits, which are typically polymers in biological cases, become persistently crosslinked or entangled, leading to a semisolid state in which flow occurs under force, In glasses, there is kinetic arrest ("frustration") into metastable energy minima. Each of these materials can maintain their character as unitary forms of matter under equilibrium or near-equilibrium conditions. This does not pertain to BMCs, which are only constituted in active, energy consuming environments. While they are not, in general, dissipative structures in the classic sense (Goldbeter, 2018; Prigogine, 1969), since some can be maintained in the absence of energy and chemical fluxes, they can sustain active processes and dynamical instabilities (and depend on them for their functional roles as in the nuclear FACs described above) when such fluxes are in effect (Lyon et al., 2021).

While the interactions within classic forms of soft matter are isotropic and chemically nonspecific, this is also not the case for BMCs, which are built from sequence-encoded macromolecules – mainly protein and RNA molecules, with varying constraints on specificity of their sequences, but none of these as stringent as in genetic mechanisms. Some BMC-resident molecular species undergo multivalent, heterogeneous interactions in continuously reconfigurable networks (Galvanetto et al., 2025). Further, BMCs exhibit a wide variety of mechanochemical properties. For example, tunable spectrums of mechanical relaxion times mak them simultaneously liquid-like and solid-like for different process. Further, chemically-driven reshaping of free-energy landscapes and reversible, regulated maturation of material properties also distinguish BMCs from classical soft matter, excitable media, and self-organizing systems (Lyon et al., 2021; Sanfeliu-Cerdán and Krieg, 2025). In light of their often *non-equilibrium dynamics* (e.g., ATP-dependence of some functions), *kinetic trapping* (i.e., failure to sample all accessible microstates freely), and *compositional heterogeneity* (i.e., different molecular species existing in different physical states), analysis in terms of thermodynamic functions of state is either ambiguous or inapplicable for BMCs. The degree of complexity that a biological system constructed of components heavily dependent on BMCs for its function, such as nervous systems and brains (Hoffmann et al., 2025), must attain to reach a "thermodynamic limit," that is the asymptotic regime at which these exotic effects average out, is an open question.

## 7. Conclusion

We have surveyed the physics of a wide range of forms of living matter. Our examples were chosen to demonstrate that the conceptual maps between these forms cannot be captured using traditional philosophical tools such as reductionism, no matter how liberally construed (Nagel 1961; Wimsatt 1976; Sarkar 1992, 1998, 2014). Reductionism tries to capture two distinct relations: that between wholes and parts and that between two domains, one of which has epistemic primacy over the other. Typically, the domain that hosts the parts has this primacy: the structure and dynamics of entities and processes of that domain are supposed to explain those of wholes. But, as we have seen, inherencies of different forms of living matter prevent any credible attribution of epistemic primacy to any single form: for instance, biogeneric processes are different from generic ones but the latter do not have epistemic primacy over the former.

The question of parts and wholes is more complicated. Biological systems have spatial organization in which most structures apparently obey straightforward mereological relationships and structural units seem to fit naturally into a unique hierarchy of levels of organization. But, once we probe the details, this view turns out to be simplistic. As Nagel (1952) pointed out in an early discussion of the parts-whole problem, how we divide an entity into parts requires theoretical choices. Should a multicellular organism be conceptualized as composed of cells, tissues, both, or other units? Since the cells of an animal body are not autonomous agents but are differentiated into specialized, functionally coordinated types (as in a kidney nephron or myelinated neuron), and social amoebae are alternately independent and cooperative over their life cycles, no theoretical decomposition of organisms into cells can be hierarchical. Further, as noted earlier, molecular reductionism, with the flow of causal specification (sometimes called information, but see Sarkar, 1996) from DNA to RNA to protein has long been popular in the biological sciences. But the new-found roles of BMCs completely upend this, with their structures and functions (many consequential for the whole organism, not just the cells they reside in) dependent on the spatial interpenetration of these molecular species and yet to be characterized physical principles. The properties we have described do not permit a unique decomposition of living activities into levels of organization. (See Wimsatt, 2007 and Batterman, 2021 for related, complementary critical views of hierarchical framing.)

What is relevant is not so much that reductionism fails but that in these contexts it provides no insight. The philosophical category of emergence advanced by traditional organicists (e.g., Pepper 1926) fares little better: each emergent form of matter may have inherencies that are its own but emergentism also presupposes an ability to organize domains with new properties in at least a roughly linear scale as we move from one level to another. This presumption is violated because of the arguments above. The same considerations carry over to discussions of holism (Smuts 1926). However, our proposals are consistent with a dialectical organicism (Levins 1998) that is similar to the project described by Gilbert and Sarkar (2000): that doctrine holds that biological explanations cannot proceed solely from the properties of fully individuated parts notwithstanding whether it is a fact that all properties of the whole are determined by the properties of the parts. Stated differently, neither the parts nor the whole has epistemic primacy for materialism to hold true.

More importantly, our examples show that biological material entities have characteristic inherencies determined by internal structure and processes which are variously expressed under different external conditions. From an ontological perspective, this observation echoes Engels' claim that there is a leap in the dynamics of matter between different levels though (given the unity of biology) these leaps may be slight and imperceptible (Engels, 1987a, 61-2). From the perspective of the physicalism/materialism of Neurath and logical empiricism, this is the thesis that a unified language for all of science is the language of everyday objects with which measurements can be made (Sarkar, 2025). Beyond that there is no program to insist on the epistemic primacy of some basic level, say generic or thus far discovered physics.

But there is no mystery about the occurrence of differences between different contexts in which any material system is investigated. In each context, when doing science, we construct theoretical models with definite goals. Levins (1966) famously distinguished between the goals of generality, realism, and precision and argued that model-building requires tradeoff between them even within a single context. Tradeoffs mean that there are approximations: some features of the system are ignored and therefore not all models are logically consistent with each other. If we move between contexts, there is even less reason to presume such consistency. So, we build models that are specific to the context at hand. There are thus purely epistemological and pragmatic reasons to expect different "laws" in different contexts (what organicists such as Bertalanffy, Polanyi, Needham, and Woodger called "level-specific" laws; Gilbert and Sarkar, 2000). The "forms of matter" perspective described here asserts that there may also be ontological bases for such domain-specific laws.

This pluralism does not imply any disunity of science, but a recognition that any sophisticated or dialectical materialism implies that no single determinant (or material constituent) of a complex entity provides a complete account of it (Anjum and Mumford, 2018). Instead of a quest for a canonical unified theoretical framework, theorizing about biological phenomena in different contexts and across the heterogeneity of living systems becomes a search for "operational coherence" (Chang, 2022) that is, establishing reliable models of phenomena that do not have incompatible experimental outcomes and are at least loosely conceptually compatible with each other. (We adopt Chang's term, though our use differs in part from his.)

Living organisms as we have characterized them here consist of countless interpenetrating and *unevenly developing* (a term used by Marxist philosophers to contest the generality of linear, stage-like change; Thatcher, 1991) forms of matter (see Table 1). The physics of life then becomes a tapestry of models for each context and temporal and spatial scale. (The simplest situation is when the underlying framework satisfies a mereological structure—but we are emphasizing that there is no such single canonical framework.) The task of gaining knowledge of such a tapestry is then one of *coordination* (a term that Chang and others have used), in analogy to providing GPS coordinates to individual users to bring together a large variety of often disparate data sources (e.g., from different satellites) and a wide variety of sophisticated technical analyses. The point is that, in any given situation, we know what we can do with the coordinates and have some idea of the limitations with respect to prediction and other uses.

Finally, we contrast this view of biological physics with the long-fashionable notions of life (and increasingly, mind) as the deployment of "information" and "computation" (Sarkar, 1996; Newman, 2024). While physical properties of cells, tissues, and gene and neuronal networks have inevitably been incorporated into these treatments, information always comes out on top of the hierarchy. Instead, we assert that the physics of living matter, not abstract representations, codes or programs, has always set the range of possibilities in physiology, development, and evolution. This pertains to the physics we partly understand like the morphodynamics of animal and plant tissues, the physics-plus-contexts – buried in history – which was responsible for the first cells, and the physics of subcellular condensed materials which we are far from understanding but can study in action, and which might have had a role in life's origin.

**Further Reading**

Table 1 Categories and examples of matter discussed in this paper

| Category | Defining properties | Examples: Nonbiological | Examples: Biological |
|---|---|---|---|
| Nonunitary | Internally structured, nonhomogeneous | Atoms, molecules | Cells; viruses |
| Generic unitary | Homogeneous above subunit scale | Parcels of chemical elements; uniform liquids; solids | Mucus; synovial fluid, tree sap |
| Biogeneric unitary | Cellular subunits, homogeneous above subunit scale, generic-type physics | N/A | Epithelioid and mesenchymal animal tissues |
| Non-generic unitary | Cellular subunits, homogeneous above subunit scale, nongeneric physics | Composite materials; alloys | Plant tissues, fungal tissues, aggregated starfish embryos, (brain tissue?) |
| Non-generic unknown | Internal structure, inherencies and physical principles all unknown | No known examples | Biomolecular condensates |